\documentclass[authorversion ,sigconf]{acmart} 

\AtBeginDocument{%
  \providecommand\BibTeX{{%
    \normalfont B\kern-0.5em{\scshape i\kern-0.25em b}\kern-0.8em\TeX}}}

\copyrightyear{2023} 
\acmYear{2023} 
\setcopyright{acmlicensed}\acmConference[UIST '23]{The 36th Annual ACM Symposium on User Interface Software and Technology}{October 29-November 1, 2023}{San Francisco, CA, USA}
\acmBooktitle{The 36th Annual ACM Symposium on User Interface Software and Technology (UIST '23), October 29-November 1, 2023, San Francisco, CA, USA}
\acmPrice{15.00}
\acmDOI{10.1145/3586183.3606764}
\acmISBN{979-8-4007-0132-0/23/10}

\usepackage{ccicons}
\usepackage{xspace}
\usepackage{url}

\newcommand{\etal}{et al.\@\xspace} % Prints ``et al.'' with proper spacing
\newcommand{\eg}{e.g.\@\xspace} % Prints ``e.g.'' with proper spacing
\newcommand{\commentout}[1]{}

\usepackage{color}
\usepackage{setspace}
\definecolor{Orange}{rgb}{1,0.5,0}
\definecolor{DarkGreen}{rgb}{0,0.5,0}
\definecolor{Purple}{rgb}{0.7,0,0.7}
\definecolor{Blue}{rgb}{0.2,0.2,0.8}
\definecolor{Red}{rgb}{1.0,0.0,0.0}
\definecolor{Brown}{rgb}{0.7,0.4,0.1}

\usepackage{multicol}
\usepackage{booktabs}
\usepackage{enumitem}

\usepackage{array}
\usepackage{longtable}
\begin{document}

\title[Telextiles: End-to-end Remote Transmission of Fabric Tactile Sensation]{Telextiles: End-to-end Remote Transmission of\\ Fabric Tactile Sensation}


\author{Takekazu Kitagishi}
\email{kitagishi-takekazu588@g.ecc.u-tokyo.ac.jp}
\orcid{0000-0002-5370-8841}
\affiliation{%
  \institution{The University of Tokyo}
  \city{Tokyo}
  \country{Japan}
}

\author{Yuichi Hiroi}
\email{yuichi.hiroi.1@gmail.com}
\orcid{0000-0001-8567-6947}
\affiliation{%
  \institution{The University of Tokyo}
  \city{Tokyo}
  \country{Japan}
}

\author{Yuna Watanabe}
\email{yuna-watanabe1923@g.ecc.u-tokyo.ac.jp}
\orcid{0000-0001-7530-8788}
\affiliation{%
  \institution{The University of Tokyo}
  \city{Tokyo}
  \country{Japan}
}

\author{Yuta Itoh}
\email{yuta.itoh@iii.u-tokyo.ac.jp}
\orcid{0000-0002-5901-797X}
\affiliation{%
  \institution{The University of Tokyo}
  \city{Tokyo}
  \country{Japan}
}

\author{Jun Rekimoto}
\email{rekimoto@acm.org}
\orcid{0000-0002-3629-2514}
\affiliation{%
  \institution{The University of Tokyo}
  \institution{Sony CSL Kyoto}
  \city{Tokyo / Kyoto}
  \country{Japan}
}

% \renewcommand{\shortauthors}{Takekazu Kitagishi, Yuichi Hiroi, Yuta Itoh, and Jun Rekimoto}
\renewcommand{\shortauthors}{Takekazu Kitagishi, Yuichi Hiroi, Yuna Watanabe, Yuta Itoh, and Jun Rekimoto}

\begin{abstract}

The tactile sensation of textiles is critical in determining the comfort of clothing. For remote use, such as online shopping, users cannot physically touch the textile of clothes, making it difficult to evaluate its tactile sensation. Tactile sensing and actuation devices are required to transmit the tactile sensation of textiles. The sensing device needs to recognize different garments, even with hand-held sensors. In addition, the existing actuation device can only present a limited number of known patterns and cannot transmit unknown tactile sensations of textiles.
To address these issues, we propose Telextiles, an interface that can remotely transmit tactile sensations of textiles by creating a latent space that reflects the proximity of textiles through contrastive self-supervised learning. We confirm that textiles with similar tactile features are located close to each other in the latent space through a two-dimensional plot.
We then compress the latent features for known textile samples into the 1D distance and apply the 16 textile samples to the rollers in the order of the distance. The roller is rotated to select the textile with the closest feature if an unknown textile is detected.
\end{abstract}

\ccsdesc[500]{Human-centered computing~Haptic devices}
\ccsdesc[500]{Computing methodologies~Learning latent representations}
\ccsdesc[300]{Computing methodologies~Neural networks}
\ccsdesc[300]{Human-centered computing~User studies}
\ccsdesc[100]{Computing methodologies~Dimensionality reduction and manifold learning}
\ccsdesc[100]{Human-centered computing~Virtual reality}
\ccsdesc[100]{Human-centered computing~Mixed / augmented reality}

% \keywords{Tactile perception; Texture; Texture recognition; Texture perception; Machine learning; Self supervised learning; Haptic feedback; Passive haptic feedback; Tactile Display}
% \keywords{Tactile perception; Texture; Texture recognition; Machine learning; Self supervised learning; Haptic feedback}
\keywords{Tactile perception; Texture; Texture recognition; Texture perception; Machine learning; Self supervised learning; Haptic feedback; Passive haptic feedback; Tactile Display}

%% A "teaser" image appears between the author and the affiliation
\begin{teaserfigure}
\centering
  \includegraphics[width=\textwidth]{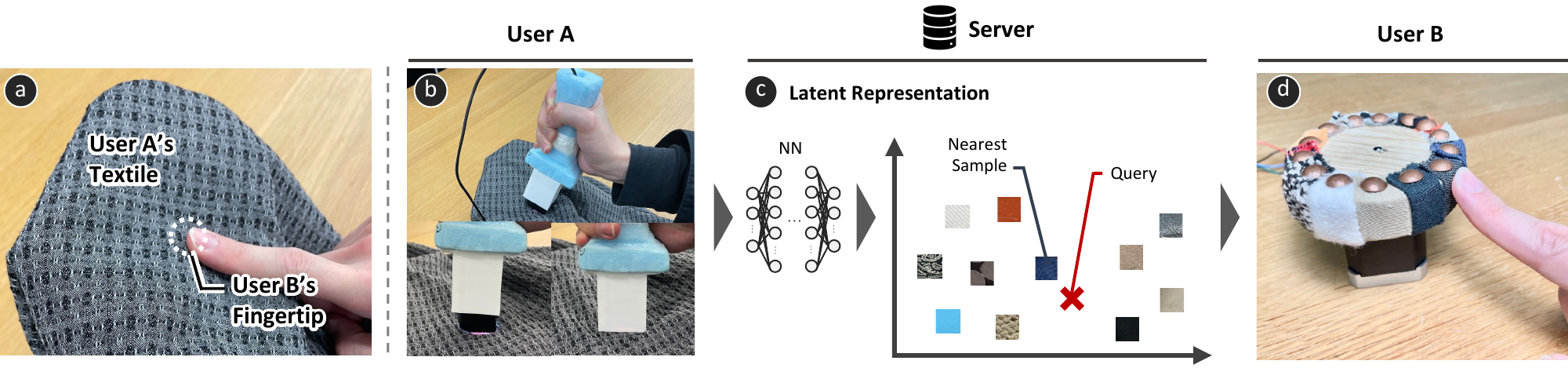}
  \caption{Overview of our Telextiles system. (a) Our ideal of remote tactile transmission. Our Telextiles aims to reproduce the tactile sensation of textiles at a remote location as if they were being touched there. 
  (b) A user transmits the tactile sensation of the textile to the server using tactile sensors. (c) On the server, the latent space is learned in advance from multiple textiles, taking into account their tactile characteristics. The user's data is converted into a latent feature, and the closest textile sample is selected. (d) The other remote user touches a physical sample of the selected textile on the latent space via an actuator.}
\label{fig:teaser}
\vspace{2mm}
\end{teaserfigure}

\maketitle

%%%%%%%%% introduction
\section{Introduction}
The tactile sensation of textiles is critical to an individual's clothing experience. In a physical store, customers often rely on touch to evaluate the texture of the textile. This tactile experience significantly shapes their evaluation of garment attributes, including comfort and quality, and ultimately impacts their overall satisfaction. However, when shopping online, customers cannot physically examine the feel of the textiles, making it difficult for them to accurately anticipate the comfort level of the garment. This limitation can reduce the quality of the shopping experience, as the garments purchased may not meet the customer's comfort expectations.

To address the challenges, we seek to develop a system that can transmit the tactile sensation of textiles to a remote location, allowing users to feel as if their fingers have touched the textile on-site. Such transmission requires both tactile sensing and actuation devices.

The sensing device must be able to detect various textiles, even with unstable hand-held sensors. An example of a simple tactile sensor is an elastomer sensor~\cite{Takahashi2019-zi, Degraen2021-rg} used for the tactile sensing of robot arms. This sensor detects the surface texture of an object by sensing its unevenness when pressed. However, the pressing force and orientation depend on the person, leading to unstable inference. In addition, existing texture recognition methods mainly classify only known textile types~\cite{li2013sensing, 8793967, s21155224, pestell2022artificial}, making it difficult to determine the relationship between known and unknown textiles that are not in the training data.

To reproduce the sensation of textiles, we ultimately want an actuator that can generate all the tactile sensations. However, it is currently difficult to reproduce the tactile sensation of textiles. 
Haptic Revolver~\cite{Whitmire2018-zn} can present a finite number of physical textures on an actuated wheel to physically convey the tactile sensation. However, it is difficult to select an appropriate finite pattern from a large number of textiles. In addition, it is an open question about which pattern will be presented when a person touches a textile that is not included in the patterns.

We propose Telextiles, an interface that enables remote transmission of tactile sensations of textiles, even when the input textile is unknown. Our system uses a sensing device that uses contrastive self-supervised learning to create a latent space that reflects the proximity of textiles. 
We also create an actuation device with 16 textile samples attached according to the proximity obtained by dimensional reduction of the latent features. 
When a remote user perceives the texture of the textile with a tactile sensor, the device is actuated to touch the physical sample closest to the user's fingertips.
Since our Telextiles interface can theoretically accept any textile as input, users can simulate the tactile experience of any garment online, which is expected to enhance the user's shopping experience and facilitate remote collaboration in the fashion and textile industries.

Our main contributions include the following:
\begin{itemize}
    \item Latent representations based on self-supervised contrastive learning considering the proximity of textiles,
    \item Apparatus for user-independent, stable tactile sensing of textiles, and
    \item A roller-type actuation device that presents a textile that is in the closest proximity to the transmitted textile sensation.
\end{itemize}

\begin{figure*}[tb]
    \centering
    \includegraphics[width=\linewidth]{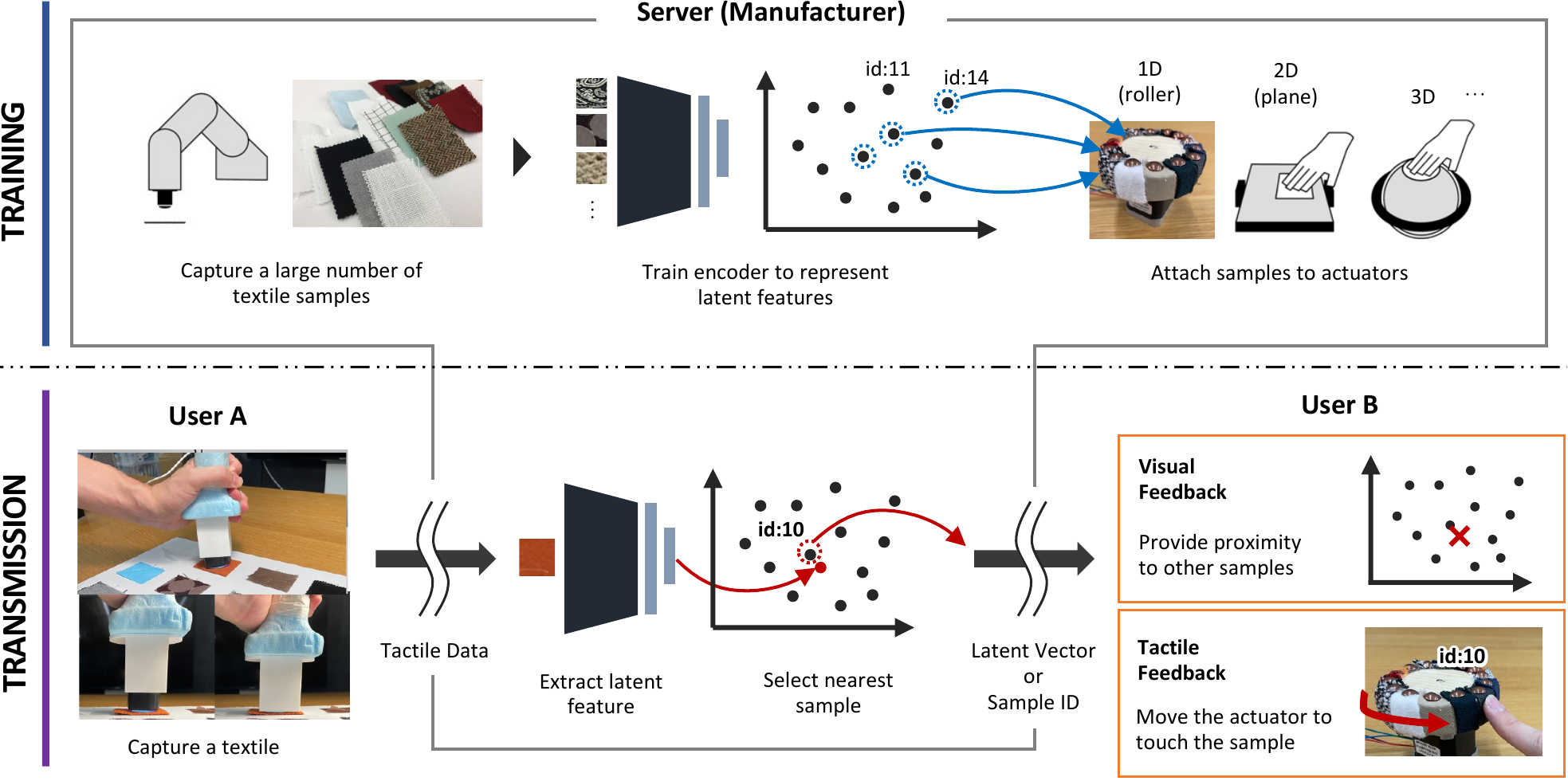}
    \caption{System overview of Telextiles. The training phase is assumed to be performed by the manufacturer, who trains encoders to extract a latent vector representation of the tactile sensation from a large sample of textiles. In the transmission phase, user A acquires tactile data by pressing a tactile sensor against a textile. This tactile data is sent over the network to the server. Using the learned encoder, the server extracts latent representations from the transmitted tactile data. The server then computes the closest training sample to the transmitted data in latent space and sends the raw latent vector or the ID of the closest sample to User B. User B can confirm the received tactile sensation through visual or tactile feedback.
    }~\label{fig:system-overview}
\end{figure*}
\vspace{-1mm}

%%%%%% related work
\section{Related Work}
In this section, we review related work on tactile recognition approaches and actuators as components for remotely transmitting the tactile sensation of textiles.

\subsection{Textile Recognition}
Research on textile recognition is mainly aimed at helping robots to understand the sense of touch. Therefore, textile recognition often proposes a combination of a tactile sensor attached to the robot and a learning model to recognize its texture. Due to this motivation, most of the related research focuses on solving the classification problem of which textile the robot is touching.
Examples of tactile sensors include robot finger friction~\cite{8793967}, a bionic tactile sensor~\cite{s21155224}, point cloud deformation on a silicone rubber~\cite{pestell2022artificial}, and an elastomer sensor such as GelSight (GelSight, Inc.)~\cite{li2013sensing}.
With these supervised classifications, it is difficult to present a plausible tactile textile when an unknown textile is used as input.

As an example of supervised learning applied to understanding unseen textiles, Yuan~\etal~\cite{8461164} combined GelSight and a 3D camera to classify the physical properties of fabrics (\eg, thickness, fuzziness, softness), which were set to five levels based on subjective evaluation. 
This method is based on a subjective rating scale and does not allow for a comparison of how close an unknown textile is to an existing one.

Several methods have been proposed to learn latent representations of textiles and apply them to robot controls. Takahashi~\etal~\cite{Takahashi2019-zi} used unsupervised learning of images from uSkin sensor (XELA Robotics) to estimate the robot's end-effector signals from latent features. Conversely, Narang~\etal~\cite{Narang2021-re} used latent features obtained by semi-supervised learning with simulations to estimate the 3D deformation of BioTac sensor (SynTouch Inc.) from the robot's end-effector signals. Although these methods use latent features to convert between different modalities, the latent features are not learned to reflect the proximity between textiles. 

To convert data between different tactile sensors, BioTac and RoboSkin sensor (Bibop Inc.), Gao~\etal~\cite{Gao2021-rl} learned a common latent space between the sensors using self-supervised learning with a two-stage recurrent network. Although their method designs a loss function that classifies the physical properties of the cloth from the latent features, these physical properties are also set based on subjective evaluations, similar to~\cite{8461164}.
We instead use self-supervised contrastive loss functions for training, so that textiles with similar characteristics are closer together and textiles with different characteristics are further apart.

\vspace{-1mm}
\subsection{Tactile Actuators to Reproduce Textiles}
Our goal is a system that presents tactile sensations as if one's own fingers were remotely present and tracing the surface of a textile. To present such tactile sensations, an actuator is required that allows the user to feel contact with the surface and explore its texture using the skin of the fingertip. Traditional handheld VR controllers are limited to vibrotactile stimulation and are not suitable for presenting these fine tactile sensations of textiles.

Examples of devices that provide tactile sensations on the skin at the fingertips include glove exoskeletons~\cite{Massie1994-wu, Bouzit2002-ta, Fang2020-fi}, finger-worn haptic devices~\cite{Prattichizzo2013-xj, Pacchierotti2017-zo, Chinello2015-af}, robot-based solutions~\cite{Araujo2016-ed, Vonach2017-sa, Suzuki2019-cc, Gonzalez2020-lv, Suzuki2021-ym}, and ultrasound~\cite{Watanabe1995-sf, Wiertlewski2015-cb}. Glove exoskeletons and finger-worn haptic devices are unsuitable for our applications where users explore the tactile qualities of surfaces because the devices are worn on the fingertip. 
Robot-based and ultrasound solutions require expensive or large setups and therefore do not match our motivation to easily experience the tactile sensation of textiles while shopping online. Therefore, we are looking for an actuator with a relatively simple setup that does not require a device on the fingertip.

Methods have been proposed to reproduce the delicate tactile sensation of textiles by physically replicating the microstructure of the textile surface, such as 3D printing based on the hair structure~\cite{Degraen2019-vv} or the digitized reconstruction of the elastomeric sensor~\cite{Degraen2021-rg}. These methods have reproduced textile tactile sensations at very high resolution as static materials. On the other hand, our system requires an actuator that dynamically selects the material with the closest tactile response to the input from multiple materials.

Wheel-based haptic controllers are proposed that convey the tactile properties of fabrics by attaching fabric samples to the actuator surface. HapticRevolver~\cite{10.1145/3173574.3173660} is an actuator wheel that rises under the finger and makes contact with a virtual surface. Haptic Palette~\cite{10.1145/3393914.3395870} extends this wheel-based actuator with visual enhancements on top of the physical texture in virtual reality environments, allowing the user to experience mixed material perception. In these wheel-based actuators, multiple physical textures are attached to a wheel that rotates to bring the optimal physical texture into contact with the fingertips. While these wheel-based actuators inspire us, they do not provide a logic to present the closest tactile sensation when an unknown pattern is an input. We address this problem by learning the distance on the latent representation with respect to the tactile sensation of the textile.

%%%%%%%%%

\begin{figure*}[tb]
    \centering
      \includegraphics[width=\linewidth]{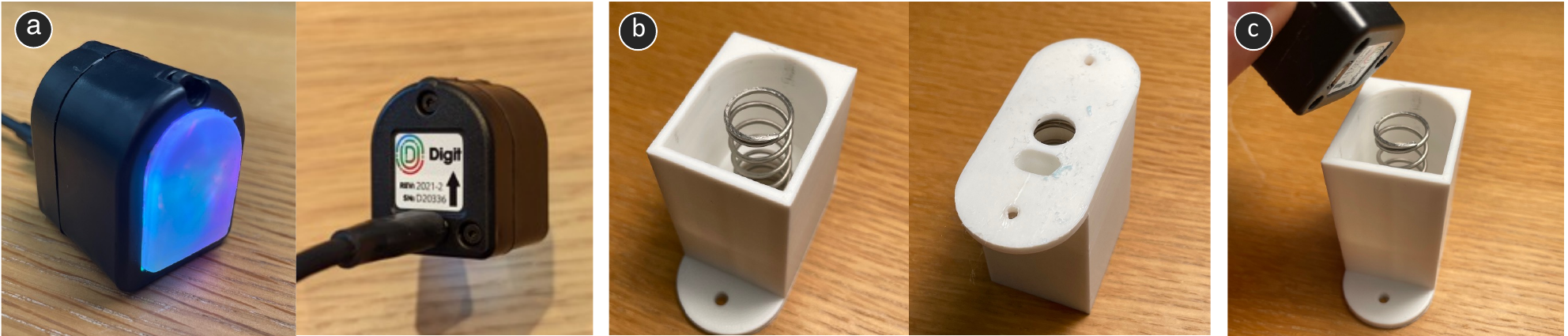}
      \caption{(a) The DIGIT optical tactile sensor. (b) The jig structure. The jig features a hole for inserting the spring and another hole for accommodating the sensor attachment cord. (c) Inserting the sensor into the jig. The figure demonstrates the custom-made jig tailored to the sensor's dimensions and the incorporated spring mechanism for applying pressure to the sensor.}~\label{fig:sensor_device}
\end{figure*}

%%%%%%%%%% proposal
\section{System Overview}
Figure~\ref{fig:system-overview} shows an overview of our Telextiles system. 
Our system can be divided into two phases: a training phase, which is performed in advance, and a tactile transmission phase, in which users actually transmit the tactile sensation of the textile. 

The training phase is assumed to be performed by the manufacturer, who trains encoders to extract a latent vector representation of the tactile sensation from a large sample of textiles.

The transmission phase is performed by two remote users, A and B. User A acquires tactile data by pressing a tactile sensor against a textile. This tactile data is sent over the network to the server. Using the learned encoder, the server extracts latent representations from the transmitted tactile data. The server then computes the closest training sample to the transmitted data in latent space and sends the raw latent vector or the ID of the closest sample to User B. User B can confirm the received tactile sensation through visual or tactile feedback. With this sequence of events, Telextile enables end-to-end remote transmission of the tactile sensation of textiles.

The following three components are required to realize the Telextiles system.
From the next section, we describes these three components, including implementation details:

\begin{enumerate}
\item Stable sensing device (Sec.~\ref{subsec:impl-sensing}): Since tactile sensors are typically attached to the robot arm, variations in how the user measures have not been considered. We implement a jig with a mechanism that keeps the force on the sensor constant and stabilizes the contact angle with the textile.
\item Learning good latent representation (Sec.~\ref{subsec:impl-encoder}): The latent space that the encoder learns should implicitly learn the proximity between the textures of each textile, \eg smoothness, stiffness, softness, etc. We use contrastive self-supervised learning for the encoder to obtain a latent space that accounts for the proximity between tactile sensations of textiles.
\item Tactile reproducibility (Sec.~\ref{subsec:impl-feedback}): There is a need for a tactile device that can well reproduce the transmitted tactile sensation. We implement a roller-type actuator that can present the physical textile with the closest tactile feel to the target textile.
\end{enumerate}

%%%%%%%%%%

%%%%%%%%%% implementation
\section{Tactile Sensing Device}\label{subsec:impl-sensing}
Figure~\ref{fig:sensor_device} shows our tactile sensing device. This device consists of an optical tactile sensor, DIGIT \cite{9018215} by Facebook AI, with a custom jig.
The combination of the custom jig and the DIGIT sensor allows for consistent and accurate acquisition of tactile information from various textiles. 
    
\begin{figure}
\centering
    \includegraphics[width=\columnwidth]{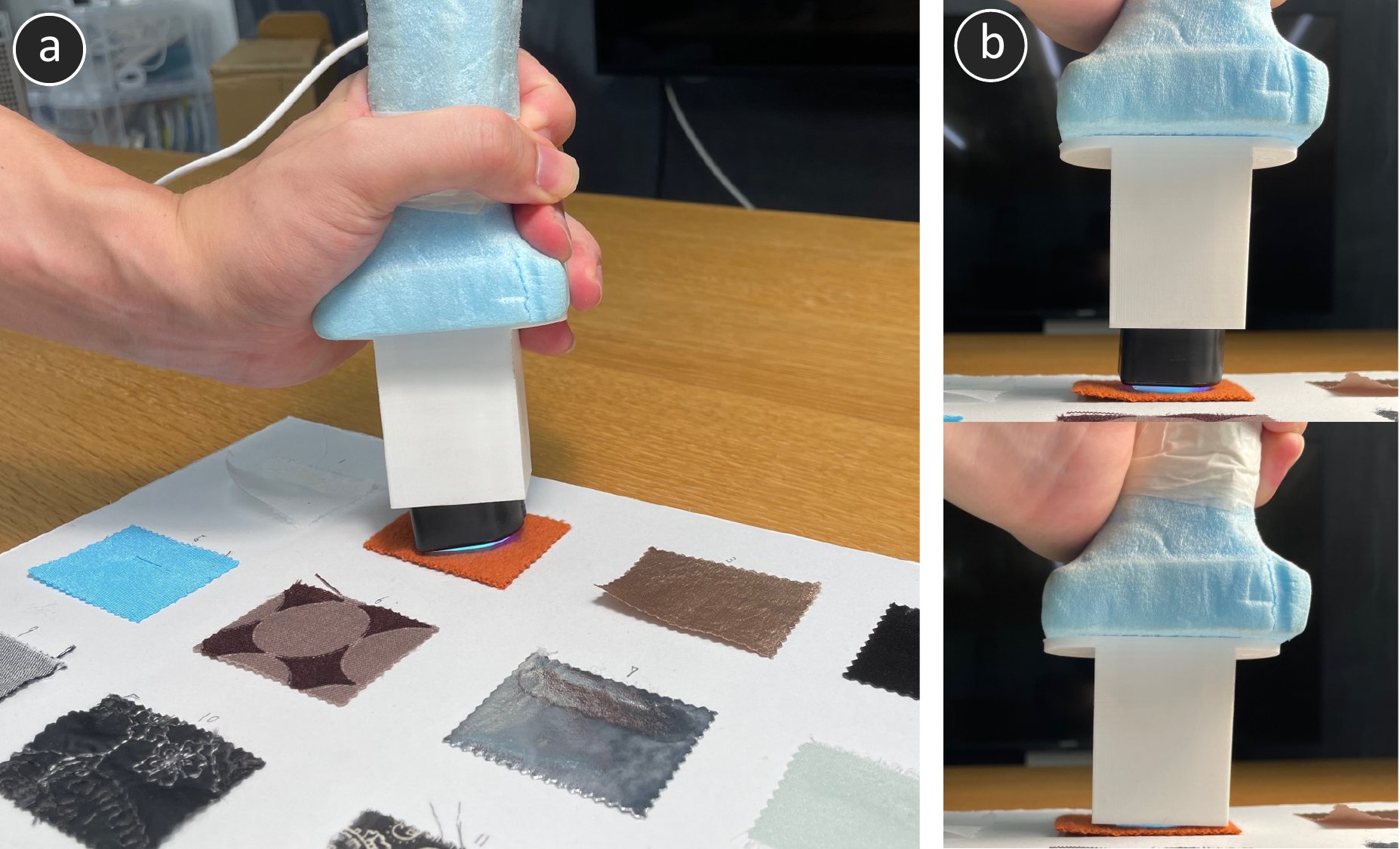}
    \caption{(a) The user capturing the tactility of textiles with our tactile sensing device. (b) The sensor is being pressed against the textile. It can be confirmed that the sensor is pressed in with a constant force due to the spring force.}~\label{fig:sensor-how-to-use}
\end{figure}
        
\subsection{Tactile Sensor}
The DIGIT sensor is a vision-based tactile sensor. It consists of a deformable reflective elastomer, RGB LED lights, and a camera (240 $\times$ 320 pixels). When pressed against a complex surface, such as a fabric, the elastomer deforms according to the texture, the scattered light illuminates the deformation, and the camera captures the surface deformation, which in turn records the microstructure of the textile surface, providing detailed tactile information. Our system used this imaging modality for input and created a latent space using self-supervised contrastive learning. 

Our decision to use the proposed approach rather than other texture-sensing sensors was largely based on resolution. Considering the case of online shopping, a high resolution that captures the nuanced details of a textile's texture, which a photo can't fully capture, is important, and in this regard, DIGIT has an advantage over other modalities, including acoustics (microphones, ultrasonic, etc.) and friction. The DIGIT sensor is also known for its lightweight, compact design, which was also suitable for the online shopping scenario.

\subsection{Attached Jig}
The jig is a device for accurate and consistent data collection. It is designed to accommodate the dimensions of the sensor and maintain a constant angle and force when pressed against the textile. Considering that the force and direction of the pressure varies from person to person, it allows different textiles to be detected, even with unstable hand-held sensors.
The jig consists of the custom 3D printed component and a built-in spring mechanism. The custom 3D printed component has 2mm of clearance on all sides to ensure a tight fit between the DIGIT sensor and the jig to reduce any potential shifting or movement. It also has a hole for inserting a spring and another for threading a cord. 
The built-in spring mechanism functions to apply a constant force to the sensor, which is determined by the spring coefficient and the dimensions of the jig. When the force exceeds a certain level, the jig stops pressing against the sensor, ensuring that the force applied is sufficient to capture the surface texture of the textile without causing damage.

Figure~\ref{fig:sensor-how-to-use} shows a user actually using the device to capture the tactile image. A Styrofoam handle is attached to the top of the jig to help the user apply vertical force.
This figure demonstrates how the jig maintains full contact with the textile surface while applying a consistent angle and pressure to the sensor, ensuring sufficient force to capture the textile's surface structure.
The 3D printing data for jig is available: \url{https://drive.google.com/file/d/1N97winpXh05WP8o95XPEtPgdJ1QJu61o/view?usp=sharing}

\commentout{The sensing device must be able to detect various textiles, even with unstable hand-held sensors, considering that the force and orientation of the pressure varies with the person, resulting in unstable conclusions.
The jig holds the sensor securely and maintains a constant angle and force as it is pressed against the textile to collect tactile data. 
The jig consists of the custom 3D printed component. It is designed to accommodate the dimensions of the sensor and maintain a constant angle and force as it is pressed against the textile, allowing for more accurate and consistent data collection. It has 2 mm of clearance on all sides to ensure a tight fit between the DIGIT sensor and the jig to reduce any potential shifting or movement. 
The jig has a built-in spring mechanism that applies a constant force to the sensor, which is determined by the spring coefficient and the dimensions of the jig. When the force exceeds a certain level, the jig stops pressing against the sensor, ensuring that the force applied is sufficient to capture the surface texture of the textile without causing damage.
It has one hole for inserting a spring and another for threading a cord.}

\section{Encoder for Tactile Representation of Textiles}\label{subsec:impl-encoder}
To process and model the tactile features of textiles, the encoder is trained with self-supervised learning. Especially, we use contrastive learning, a type of self-supervised learning, to reflect the proximity to the tactility of the textile in the latent space. In this subsection, we first describe these training methods, then the training configuration of the encoder, and the processing on the actual test data.

\subsection{Self-supervised Learning}
Self-supervised learning is a type of machine learning in which an algorithm learns to predict certain features or properties of data without explicit supervision or labels. 
By creating a latent space of tactile information for textiles, it allows us to identify relationships between different textiles, including those not present in the training data. This capability enables the system to display textiles with similar properties, even if they were not part of the original training dataset.

\subsection{Contrastive Learning}
Contrastive learning  learns to distinguish between similar and dissimilar pairs of samples. 
In contrastive learning, the model is trained on a set of data pairs, where each pair consists of two similar or dissimilar samples. The model is trained to maximize the similarity score of the similar pairs and minimize the similarity score of the dissimilar pairs. This is usually done by using a contrastive loss function.

When the encoder is trained on a set of pairs of tactile images of textiles, two images of textiles with similar tactile properties are learned to move closer together in latent space. On the other hand, two images of textiles with different tactile properties are learned to move further apart in latent space.

We employed MoCo~\cite{He2019-aq} as the contrastive loss. As far as using the contrastive loss with negative examples explained earlier, we obtained almost the same accuracy with different loss choices, such as SimCLR~\cite{Chen2020-xj}.
On the other hand, when we applied contrastive learning to DIGIT images, we found that contrast loss did not progress at all without negative examples (\eg, SimSiam~\cite{Chen2020-fx}, DINO~\cite{Caron2021-hs}, etc.). These methods are trained on pairs of two images of the same image transformed with different data augmentations. Therefore, when contrastive learning DIGIT images with these methods, it is necessary to find a data augmentation method that matches the characteristics of the DIGIT images.

\subsection{Training Configuration}
We implemented the encoder using Lightly~\cite{susmelj2020lightly}, a self-supervised learning library running on PyTorch Lightning. We will describe the detail of the model implementation and training.

\subsubsection{Network Architecture}
We used the ResNet-18 architecture as the backbone for our model. Before inputting the DIGIT image into the ResNet, the image was cropped to a size of 224 $\times$ 224 based on the center. To adapt this network for our specific task, we modified the final layer of ResNet-18 to output a 512-channel tensor. We achieved this by combining the later stages of the network with a $1\times1$ adaptive average pooling layer. Finally, we obtained a 512-dimensional feature vector.

\subsubsection{Training Data}
We assigned 11 participants (6 males and 5 females) to collect data from all $119$ samples in the book~\cite{Tanaka2009-vp} and used it for training. This dataset is a collection of images acquired using the DIGIT sensor.
Participants were given simple instructions: to press the sensor against each fabric sample until the jig made contact with the sample. We did not impose any additional restrictions or specific guidelines on how to use the sensor to accommodate individual handling variations, thereby fostering more generalized learning of the self-supervised model. As a result, some participants rotated the sensor by $90$ or $180$ degrees, and the forces that participants applied to the device varied.
For each fabric sample, images were captured at a rate of $60$ frames per second for a duration of $2.5$ seconds, resulting in a total of $150$ images per sample.

\subsubsection{Data Augmentation}
To improve the robustness of our model, we employed data augmentation techniques. We retained most of the settings from the general contrastive training model, SimCLR~\cite{Chen2020-xj}. However, we made some adjustments to better suit the characteristics of the DIGIT sensor images.

First, because the DIGIT image detects microstructures from three colors of LED light, we turned off data enhancements related to hue change, imposition of Gaussian blur, and gray-scaling.

Next, given that the texture microstructure of the textile is invariant to rotation and mirror symmetry, we adjusted the image rotation settings: set probability that vertical flip to be 0.5, set probability that random rotaion to be 1.0, and the set random rotation range to $[-180^{\circ}, 180^{\circ})$.

Finally, we changed the normalization parameters to accommodate the unique properties of the DIGIT sensor images. We calculated the mean and standard deviation for each of the RGB values across all the training data and used these values for normalization. We set the mean as $[0.37932363, 0.4131034, 0.38336082]$ and standard deviation as $[0.11476628, 0.08604312, 0.16590593]$.

\subsubsection{Optimizer and Epochs}
For optimizing our network during training, we utilized the Stochastic Gradient Descent (SGD) algorithm. We set the learning rate parameter to $0.03$, the momentum parameter to $0.9$, and the weight decay parameter to $10^{-4}$. We trained the model at 200 epochs.

\section{Feedbacks for Transmitted Tactile Features}\label{subsec:impl-feedback}
Feedback to the user based on tactile features transmitted from the server is provided by visual and tactile feedback mechanisms.
To effectively communicate the relationships between textiles, the original high-dimensional latent features need to be reduced to lower (generally, one or two) dimensions for user-friendly visualization and interaction. 
To achieve this, UMAP (Uniform Manifold Approximation and Projection)~\cite{McInnes2018-vh} is employed for dimensionality reduction. UMAP is a non-linear dimensionality reduction technique that preserves both local and global structures in the data, which is effective in visualizing high-dimensional data in a lower-dimensional space while maintaining the relationships between data points.

With the reduced-dimensional features, Telextile can provide the user with visual feedback through an interactive 2D map, and tactile feedback through a motorized device that brings the closest matching textile to the user's hand for comparison. 
The combination of the tactile feedback and the visual feedback provided by the 2D map offers users a comprehensive way to explore textiles with similar textures, even if those textiles were not part of the original training data.

\subsection{Visual feedback}
\begin{figure}[tb]
\centering
  \includegraphics[width=0.85\columnwidth]{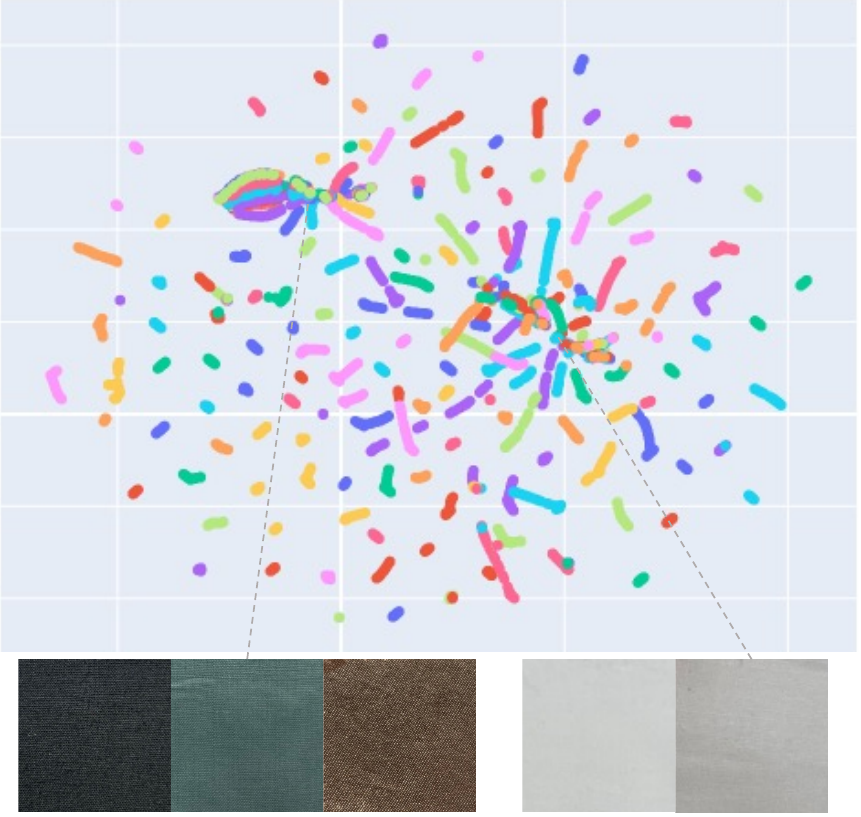}
  \caption{UMAP 2D visualization of the feature vectors of the training data (119 different textiles). It can be seen that the majority of the samples belong to separate clusters. Textiles with similar weave patterns, rather than similar tactile properties, cluster more closely in latent space, suggesting the limitations of the DIGIT sensor in capturing distinct tactile sensations. For example, the fabrics in the sets have similar weave and yarn thickness, but different materials and yarn stiffness. The three fabrics on the left, from left to right, were made of wool, cotton, and silk, with the left feeling stiffer. The two fabrics on the right, from left to right, were cotton and linen, with the left feeling stiffer.}~\label{fig:umap-2d-space}
\end{figure}

The visual feedback is designed to give users a way to visually see how close the tactile information sent is to the existing textile.
It shows the relationship between a textile touched by a remote user and a set of textiles previously registered in our system. Figure~\ref{fig:umap-2d-space} shows the mapping. Users can access this mapping when the remote user presses the sensor. 
In addition, as the user moves the cursor over the dots, additional information is displayed to assist the user in imagining the tactile feel of the fabric, which could include an image of the texture. It also has the ability to zoom by specifying an area, allowing the user to see the relationship between the dots in detail. For example, by specifying the area around the test fabric data and zooming in, the relationship between the test fabric and the already registered fabric can be seen in detail. We created this interactive 2D map by reducing the original high-dimensional feature vectors to two dimensions using UMAP and visualizing them using plotly \cite{plotly}.

In the future, this visual feedback will provide a way for users to explore and compare textiles based on tactile characteristics.
For example, when shopping online, users can explore garments that feel similar to the garment they have already purchased, or see how similar the feel of garments they are considering purchasing is to the feel of a typical textile.
In these applications, this visual feedback can be combined with tactile feedback, discussed below, to enhance the user's understanding of the tactile experience of the garment.

\subsection{Actuator Design for Tactile Feedback}
The actuator is designed to provide tactile feedback to the user. The user can physically compare the closest matching textile to the received tactile feature vector via the actuator.

We used a roller-type actuator inspired by the HapticRevolver~\cite{10.1145/3173574.3173660}.
The sample textiles were placed on a circular wooden plate, evenly distributed around the plate, and connected to a motor. The motor is programmed to rotate the plate so that the most compatible fabric comes in front of the user. The rotation of the motor is controlled by calculating the distance between the centroid of the feature vector of the query data and the centroid of the feature vector of the selected fabric in the plate. 

The size of the wooden plate we used was 8 cm in diameter and 2 cm in thickness. Due to the limitations of the plate size, 16 textiles were selected from the 119 training samples in this study. Here we describe how we selected some textiles from the training samples to select fabrics that cover different areas in the latent space; first, we used UMAP for dimensionality reduction of the features of the training samples to 1D (scalar value) to apply to the one-dimensional structure of the plate. Then, 16 of the training samples were selected from this number line so that they were equidistant from each other.

\subsubsection{Hardware Configuration}
% Hardware configuration
\begin{figure}
\centering
  \includegraphics[width=\columnwidth]{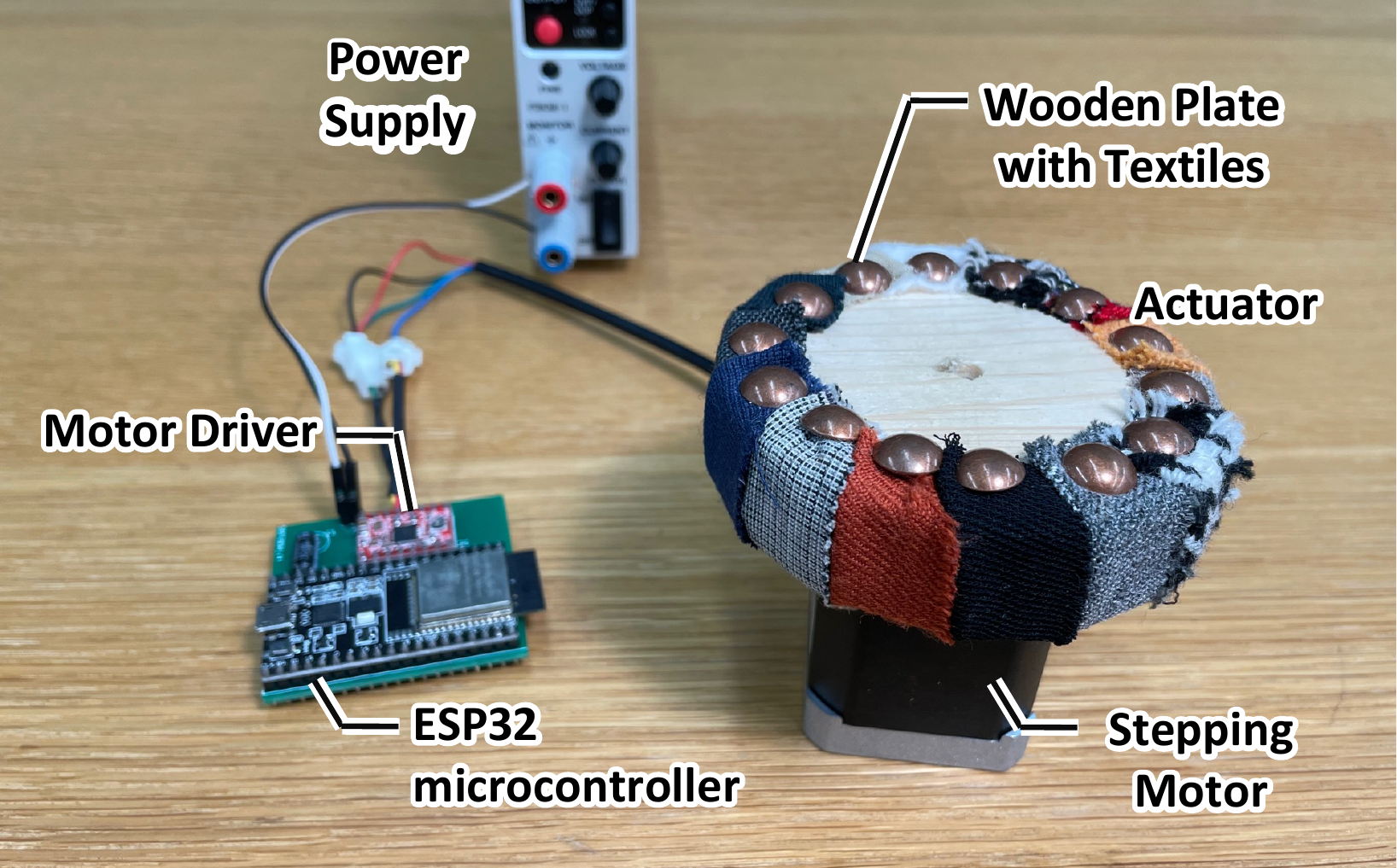}
  \caption{The hardware configuration of the actuator, including a stepping motor, a roller, a motor driver, an ESP32 microcontroller, and electronic circuitry.}~\label{fig:hardware_configuration}
\end{figure}

Figure \ref{fig:hardware_configuration} shows the hardware components of the interaction device including an ESP32 microcontroller (ESP32-WROOM-32D, Espressif System), a stepper motor (ST-42BYH1004-5013, Mercury Motor), and a motor driver (A4988, KKHMF). 

After determining the closest matching textile, the main system calculates the motor rotation angle based on the current position and target position of the plate. The ESP32 communicates with the system via Bluetooth serial and receives the calculated rotation angle. It then sends signals to the driver, and the driver controls the rotation of the stepper motor according to the received signals.

The motor rotates the plate over the shortest possible distance. The ESP32 sets the voltage on the DIR (direction) pin of the motor driver. If the rotation angle read from the Bluetooth serial is positive, the DIR pin is set to HIGH, causing the motor to rotate clockwise. If the angle is negative, the DIR pin is set to LOW, causing the motor to rotate counterclockwise.

The stepper motor has a fixed angle of rotation per step. To calculate the number of steps required to reach the target angle, the value received from the Bluetooth serial is divided by the fixed angle. The ESP32 sends pulses equal to the calculated number of steps to the STEP pin of the motor driver. The motor driver then controls the stepper motor to rotate to the desired angle.

% Software configuration
\subsubsection{Software Configuration}

The software controls the actuators by the following steps:

\begin{enumerate}
    \item Call the API every second to get a number representing the target position according to the distance in the embedding space.
    \item Calculate the angle at which the motor should turn by multiplying the number by $360 / N$, where $N$ is the number of samples on the actuator.
    \item Calculate the angle to rotate based on the stored current motor angle and the angle calculated in (2). Set the angle to be clockwise if it is closer or counterclockwise if it is closer.
    \item Send the angle calculated in (3) to the ESP32 via Bluetooth serial communication.
\end{enumerate}

%%%%%%%% evaluation
\section{evaluation}
To evaluate whether the latent space mirrors the tactile sensation of the textile, we examined how distinctly different textiles were positioned within the latent space. Then, we conducted a user test to determine whether the model could assess the similarities between fabric combinations as accurately as those perceived by individuals.
\commentout{to determine whether the tactile sensation of the textile, as selected by the actuator, mirrored the actual tactile sensation of the textile as perceived by the user.}

%%%%%%%%%%
\begin{table}
  \centering
  \begin{tabular}{l r r}
    {\small With Jig} & {\small Max@top1 (\%)} & {\small Final@top1 (\%)}\\
    \midrule
    No & 82.84 & 81.70\\
    \textbf{Yes} & 99.69 & 99.24\\
  \end{tabular}
  \vspace{3mm}
  \caption{The effect of the jig usage. By using the jig, we can improve the $k=1$ accuracy of $k$-nearest neighbor clustering. Max@top1 is the maximum accuracy during the whole training epoch, and Final@top1 is the accuracy at the end of the training epoch.}~\label{tab:table2}
\end{table}

\subsection{Latent Space Evaluation}\label{sec:eval-latent}
To examine how distinctly different textiles were positioned within the latent space, we used the $k$-nearest neighbor method to cluster the latent space and measured the accuracy by comparing these cluster labels with the actual labels.
$k$-nearest neighbor (kNN) is a machine learning algorithm that classifies a new data point by finding the $k$ nearest points in the training set and assigning the label of the majority of those neighbors to the new data point.
We trained the encoder using the training data, then determined the cluster of the feature vectors for the test data. This data was collected from a male subject in the same manner as the training data.
We applied clustering to the feature vectors and compared the results with the actual labels.

The result was that the learned latent space clustered 119 types of textiles with 80.44 \% accuracy. From the result, although the visual feedback from dimensionality compression in UMAP did not separate some data (Fig.~\ref{fig:umap-2d-space}), we confirmed that our latent space is sufficiently capable of separating textile features in higher dimensional spaces.

In addition, we evaluated the impact of using the jig on our model's capability to learn meaningful representations of fabric textures.
One participant collected data from all 119 samples in the book~\cite{Tanaka2009-vp}, with and without the jig. For each sample, the first 120 images were used for training, while the remaining 30 images were reserved for testing.
We trained the MoCo model on this data, keeping the parameters consistent with those specified in Section 5.3. We then fed the test data into the encoder to generate embeddings. These embeddings were then subjected to k-NN clustering, the results of which were compared with the actual labels.

Table~\ref{tab:table2} shows the results. The data collected with our customized jig was able to cluster with 99.69\% accuracy. In contrast, without the jig, the accuracy dropped to 82.84\%. These results confirm that the jig increases the stability of the input data.

\commentout{One participant gathered data from all 119 samples in the book referenced as \cite{Tanaka2009-vp}, utilizing the jig either with or without it. For each sample, the initial 120 images were used for training, while the last 30 images were reserved for testing. 
We trained the MoCo model on this data, maintaining the parameters identical to those outlined in the preceding section. Following this, we inputted the test data into the encoder to produce embeddings, which were then subjected to k-NN clustering to identify clusters of similar cloth textures. To evaluate the efficacy of our method, we compared the clustering outcomes with the actual fabric labels. 
Table~\ref{tab:table2} presents the results. The data collected with our customized jig was able to achieve a clustering accuracy of 99.69\%. In contrast, without the use of the jig, the accuracy dropped to 82.84\%. These findings validate that the jig enhances the stability of a user's tactile perception of textiles.}

%%%%%%%
\begin{figure}[tb]
\centering
  \includegraphics[width=\columnwidth]{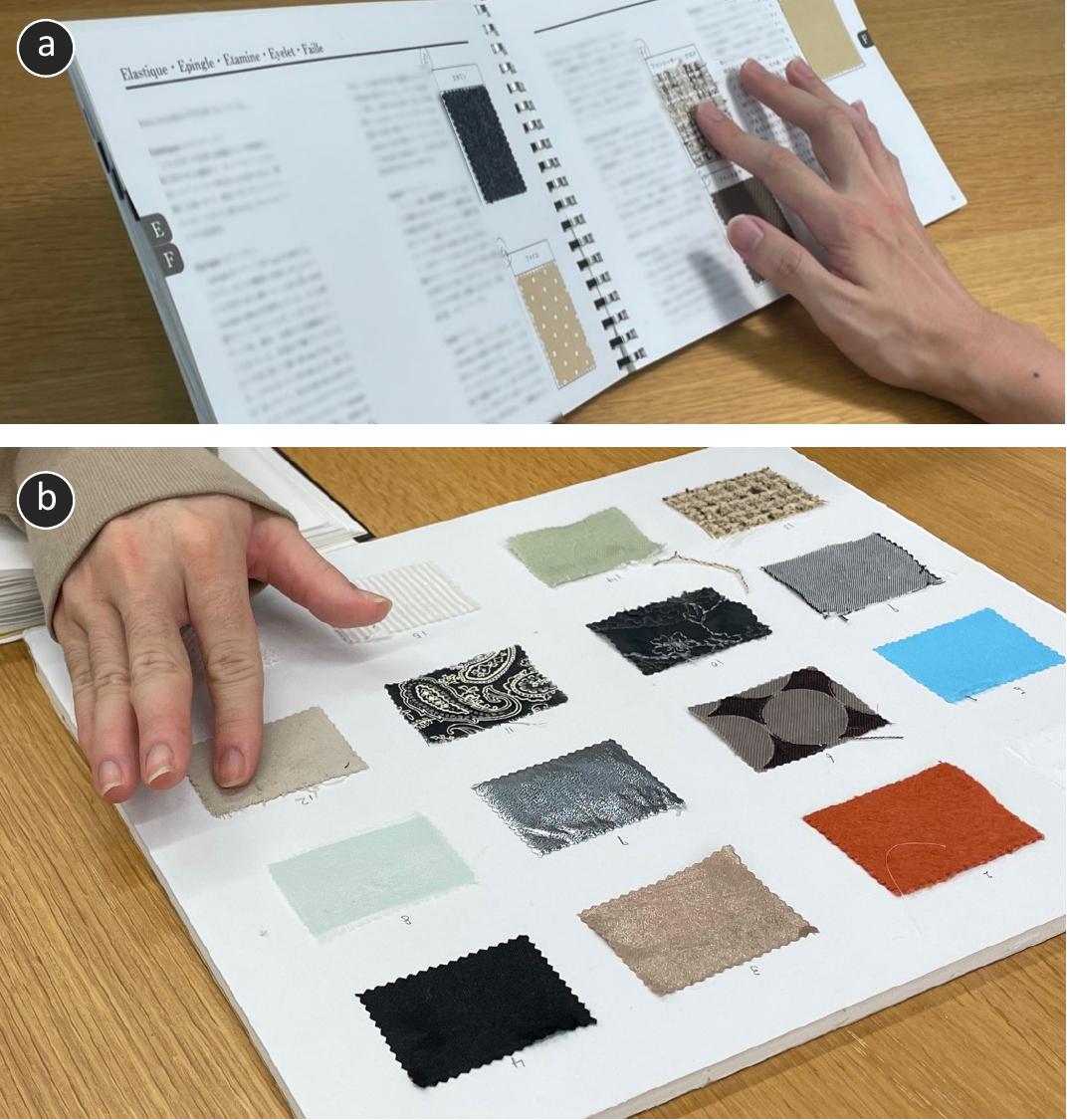}
  \caption{User Experiment. Subjects were asked to perform the following procedures three times. (a) Subjects select a pattern from [29], which includes the samples used to train our model, or [20], which includes samples not used to train the model, and touch it. (b) They then touch 16 different patterns that we have implemented on the actuator and select the five patterns that feel closest to their chosen fabric.}~\label{fig:eval-user-test}
\end{figure}

\subsection{User Test}\label{sec:eval-user}
To determine whether the model could assess the similarities between fabric combinations as accurately as those perceived by individuals, we conducted a user test.

Figure~\ref{fig:eval-user-test} shows the user test. In order to get the similarity judged by humans, the subjects performed two primary tasks. First, they were asked to randomly select one sample from \cite{Tanaka2009-vp}, which included the samples used to train our model. Second, they were asked to identify the five samples on the board, which has 16 samples we used for tactile feedback, that were most similar to the sample selected. This process was repeated three times. Subsequently, they were asked to repeat the same process with a sample from \cite{Mizushima_kakou_co_ltd2012-go}, which included samples  not used to train the model. We had eight participants—three females and five males. Two participants completed one trial each, while six participants completed two trials each, giving a total of 14.

On the other hand, we calculated the similarity judged by the model, between the randomly selected samples and the fabrics on the board. For the samples selected by the subjects in the 42 trials, we acquired tactile data using the input device. We then extracted the features in the latent space and calculated their centroids. Similarly, we calculated the centroids of the fabrics on the board. Next, to assess the similarity between the selected samples and the board samples, we calculated the Euclidean distance in the latent space between their centroids. We ordered the board samples by distance and selected the five most similar samples.

To compare the similarity judged by the human and the similarity judged by the model, we evaluated them using Top-K accuracy and a t-test for the rank correlation coefficient.
For Top-K accuracy, we calculated the proportion of 'correct' trials out of all 42 trials. We defined a trial as 'correct' if the subject judged it to be within the top 1 most similar and the model judged it to be within the top K most similar, comparing the randomly selected sample to the sample on the board.

\begin{figure}[tb]
\centering
  \includegraphics[width=\columnwidth]{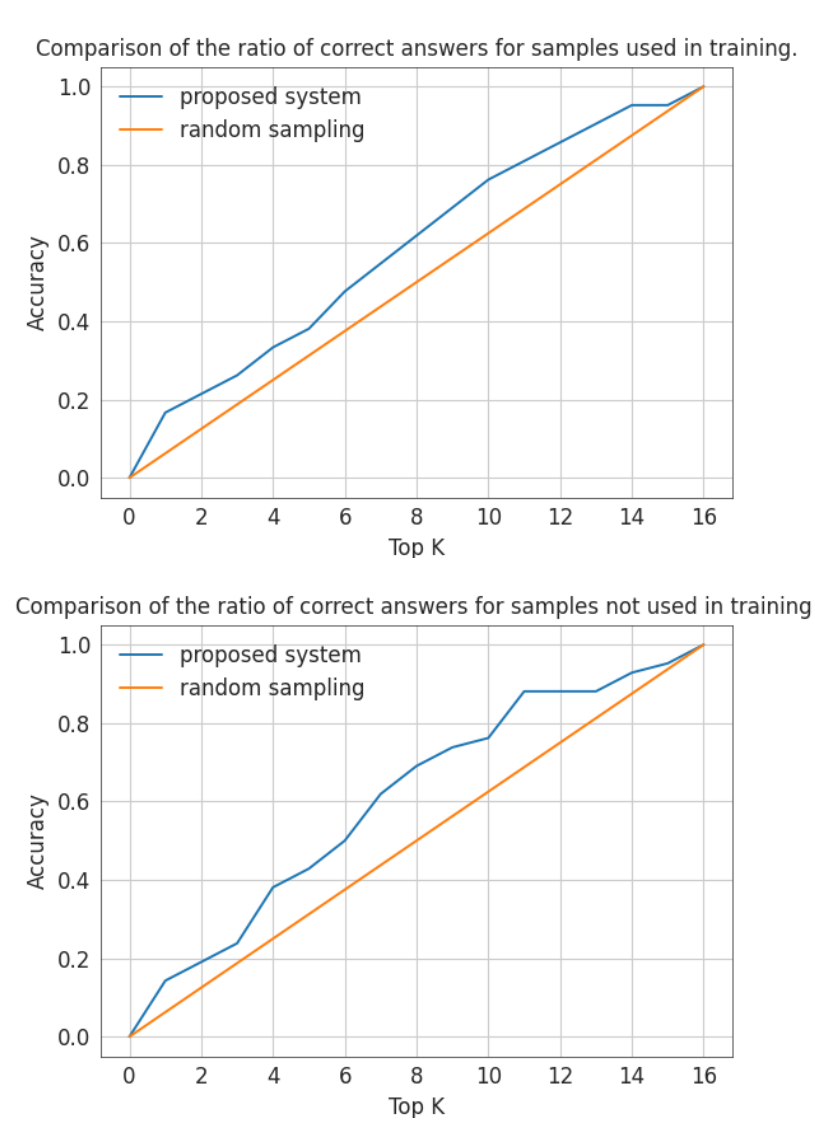}
  \caption{Comparison of human judgment and model inference in identifying the closest tactile match among 16 textiles for a randomly selected cloth sample, for samples used in training and samples not used in training. The graph presents the percentage of correct responses when the human-selected textile is within the Top K model-inferred closest textiles, compared to the expected correct answer rate from random textile choice.}~\label{fig:usertest_evaluation}
\end{figure}

Figure \ref{fig:usertest_evaluation} shows the result. We calculated the Top-K accuracy for K=0,1,2,...,16. For the samples used or not used for training, the accuracy that the encoder chose Top K for what the human chose as Top 1 was 16.7\%, 14.3\%. This indicates that the performance of the self-supervised learning model was better than random. 

For the rank correlation coefficient t-test, we compared the ranks assigned by humans and systems to see if there was a significant difference between them. We computed the Spearman rank correlation coefficient for each trial, with values close to 1 indicating a stronger correlation and values close to 0 indicating a weaker correlation. The null hypothesis of no significant difference between the correlation of each study and 0 was tested using a t-test. 

The result was that we obtained p-values of 0.518 and 0.9 for the training and non-training samples, respectively, which did not reject the null hypothesis. This means that we cannot confirm a correlation between the ranking methods of the system and those of the users.

These results suggest that the encoder's choices did not match the human subjects' choices enough, and there is room for improvement.
One possible reason for this is that humans seem to take the sense of friction into account, while the system does not. During the input phase, we simply pressed the DIGIT sensor against the surface of the cloth. On the other hand, the subjects tended to slide their fingers across the cloth.
This suggests that the pattern recognition methods used by our system and humans may not yet be fully aligned, and that sliding the sensor and capturing the sense of friction may be effective.

%%%%%%%%%%%%%%%%
%%%%%%%%%%%%%

%%%%%%%%% discussion
\section{discussion and future work}
Our system leverages the power of self-supervised learning to create a continuous latent space representation of material properties, enabling the system to adapt to unknown data and effectively match similar textures. While this adaptability allows for a more versatile and dynamic interaction with the physical world, there is still room for improvement in our system. 

The user test suggested that the judgments made by the human and the model were not in agreement. Observations during the experiment showed that when people took tactile sensations of the cloth, they slid their fingers over the cloth, not just pressed against it as the sensor did. By sliding the sensor during data acquisition, we may be able to capture a stronger sense of friction for the cloth, providing a more accurate and detailed representation of fabric textures. Additionally, incorporating a mechanism to capture time series changes in the model could improve the system's adaptability and responsiveness.

In the visual feedback system, we found that textiles with similar weave patterns, rather than similar tactile properties, clustered more closely in the latent space, suggesting the limitations of the DIGIT sensor in capturing distinct tactile sensations. Capturing additional features such as friction and thermal sensation, perhaps through sensor sliding or temperature sensing, could improve the effectiveness of the system. 

In the tactile feedback system, the number of types of tactile information that the roller can display is limited. Our choice of a roller as a feedback device is to verify the proper functioning of the latent space. Enhancing the interaction device to support 2D or 3D spatial representations of fabric tactile sensation or tactile digital fabrication, such as acoustics and microstructures, could enable our system to convey more detailed information about the fabrics, further enriching the user experience. These improvements could contribute to the development of more intuitive and user-centered interfaces in a variety of applications.

The 16 fabrics should be chosen to cover different areas of latent space. The diversity of these 16 fabrics might be better captured by ordering using the first principal component of PCA, or latent space clustering with 16 k-means. The optimal method needs further investigation.

Integrating our system with robotic platforms could automate the tactile data acquisition process, maintaining a constant force and angle during interactions. This integration could lead to more reliable and accurate results.

The versatility of the jig attachment allows for various applications and adaptations. In our study, we used a polyester handle, but the jig could be attached to different props such as sticks or boards. This flexibility expects the acquisition of tactile impressions of textiles with less force application, enabling a more consistent data acquisition process.

The miniaturization of our actuator could change the way we interact with the world, allowing for more immersive experiences in remote communication and teleoperation scenarios. On the other hand, miniaturization requires the development of research on tactile actuators that can reproduce the tactile sensation of more delicate textiles. If such miniaturized actuators are developed, such as wristwatch communication devices and fingertip sensor devices, the actuators can be expected to replace the current roller-type actuators.

%%%%%%%%%

%%%%%%%%% conclusion
\section{conclusion}
In this study we propose Telextiles, an interface that can remotely transmit tactile sensations of textiles.
Using custom tactile sensing devices, contrast learning latent space representation, and roller actuators, we demonstrated that our system is capable of end-to-end remote textile tactile sensing.
Our system is theoretically capable of mapping and presenting an infinite amount of tactile information to a finite sample, and is expected to facilitate value judgments in online shopping and telecommunications.

%%%%%%%%%

\begin{acks}
We would like to thank Hirotaka Hiraki and Naoki Kimura for their insightful discussions. We are grateful to Kiyosu Maeda for his help with server communications, and to Shota Kiuchi for his help in developing the jig. We also appreciate the contributions of all those who participated in our experiments and provided advice during lab meetings. This work was supported by JST Moonshot R\&D Grant Number JPMJMS2012, JST CREST Grant Number JPMJCR17A3, JSPS KAKENHI Grant Number JP20H05958, JST FOREST Grant Number JPMJFR206E, the commissioned research of NICT Japan, and the University of Tokyo Human Augmentation Research Initiative.

\end{acks}

\bibliographystyle{ACM-Reference-Format}
\bibliography{Source/main}

\end{document}